\documentstyle[preprint,aps,version2]{revtex}
\begin{document}
\vspace{-80ex}
\begin{flushright}
\vspace{-2.0mm}
TUIMP-TH-96/79.  August, 1996\\
\end{flushright}

\draft
\begin{title} {\bf \Large\bf $t \bar{t}$ Production Rates 
at the Tevatron and the LHC in \\Topcolor-Assisted Multiscale 
Technicolor Models }
\end{title}
\bigskip
\bigskip
\centerline{ Chong-Xing Yue$^{a,b} $,~~~~ Hong-Yi Zhou$^{a,c}$,~~~~ 
Yu-Ping Kuang$^{a,c} $,~~~~  Gong-Ru Lu $^{a,b} $ } 

%{\baselineskip=18pt
\begin{instit}
$^a$China Center of Advanced Science and Technology (World Laboratory),\\
      P. O. Box 8730, Beijing 100080, China\\
$^b$ Physics Department, Henan Normal University, Xin Xiang, 
 Henan 453002,  P. R. China \\	 
$^{c}$Institute of Modern Physics, Tsinghua University, Beijing 100084,
         China\footnote{Mailing address.}\\
\end{instit} 
%\newpage
\vspace{0.5cm}
\renewcommand{\baselinestretch}{1.0}
\centerline{ Abstract}
\vspace{-1.4cm}
\begin{abstract}
We study the contributions of the neutral pseudo Goldstone bosons 
(technipions and top-pions) to the $t\bar{t}$ production cross sections 
at the Tevatron and the LHC in topcolor-assisted multiscale technicolor 
(TOPCMTC) models via the gluon-gluon fusion process from the loop-level 
couplings between the pseudo Goldstone bosons and the gluons. The MRS
set $A'$ parton distributions are used in the calculation. It is shown 
that the new CDF datum on the $~t\bar{t}~$ production cross section gives
 constraints on the parameters in the TOPCMTC models. With 
reasonable values of the parameters in TOPCMTC models, the cross section 
at the Tevatron is larger than that predicted by the standard model, and is 
consistent with the new CDF data. The enhancement of the cross section
and the resonace peaks at the LHC are more significant, so that it is 
testable in future experiments.\\

\end{abstract}
%\vspace{4.0mm}
\pacs{PACS numbers: 12.60.Nz, 14.65.ha, 13.87.Ce}

\narrowtext
\newpage

\baselineskip=.38in
\noindent
{\bf 1. Introduction}  

 Among the yet discovered fermions, the top quark has the strongest coupling 
to the electroweak symmetry breaking (EWSB) sector. So that processes with
top quarks are good places for probing the EWSB mechanism. Experimental
measurements of the top quark mass $~m_t~$ and the $~t\bar{t}~$ production 
cross section $~\sigma_{t\bar{t}}~$ at the Fermilab Tevatron have been 
improving. In the new 1996 CDF data \cite{CDF}, $m_t=175.6\pm 5.7(stat)
\pm 7.1(syst)~$GeV and $\sigma_{t\bar{t}}=7.5^{+1.9}_{-1.6}~$pb, the error 
bars are well reduced relative to the 1995 data by the CDF and D0 
Collaborations \cite{CDFD0}\footnote{The 1996 D0 data still contains rather 
large error bars.}. The above experimental value of $~\sigma_{t\bar{t}}~$ is
slightly larger than the standard model (SM) predicted value (taking
into account of resummation of soft gluon contributions) which is around
5 pb \cite{SMsigma}. Of course, one should wait for further improved 
data to see whether this really means something. But, as the study of
the EWSB mechanism, it is interesting to study the $~t\bar{t}~$
production cross section in EWSB mechanisms other than the SM Higgs sector,
and see if the present experimental data can give constraints on the 
parameters in the EWSB models.

  Technicolor (TC) \cite{TC} is an interesting idea for naturally breaking the
electroweak gauge symmetry to give rise to the weak-boson masses. It is one of
the important candidates for the mechanism of electroweak symmetry breaking. 
Introducing extended technicolor (ETC) \cite{ETC} provides the possibility of 
generating the masses of ordinary quarks and leptons. The original ETC 
models suffer from the problem of predicting too large flavor changing neutral 
currents. It has been shown, however, that this problem can be solved 
in walking technicolor (WTC) theories \cite{WTC}. The electroweak parameter S 
in WTC models is smaller than that in the simple QCD-like ETC models and 
its deviation from the experimental central value may fall within current 
experimental bounds \cite{WTCS}. To explain the large hierarchy of the quark 
masses, multiscale WTC (MWTC) model was further proposed \cite{MWTC1}. 
However, even in this model, it is difficult to generate such a large 
top-quark mass as what is measured at the Tevatron \cite{CDF} without 
exceeding the experimental constraint on the electroweak parameter
$~T~$ \cite{CDT} even with ``strong'' ETC \cite{SETC}.
In addition, this model generates too large corrections to the
$~Z\rightarrow b\bar{b}~$ branching ratio $R_b$  compared with the LEP
data \cite{LEP} due to the smallness of the decay constant $F_Q$, and a 
consistent value of $R_b$ can be obtained \cite{YKL}\footnote{It has been 
shown that ETC models without exact custodial symmetry may give rise to 
consistent values of $R_b$ \cite{ETCRb}, but such models may make the 
electroweak parameter $T$ too large.} by combining this model with the 
topcolor interactions for the third generation quarks \cite{TOPC1} at the 
energy scale of about 1 TeV. Similar to QCD, topcolor-assisted multiscale 
technicolor (TOPCMTC) theory predicts certain pseudo Goldstone bosons (PGB's) 
including technipions and top-pions \cite{MWTC1}\cite{MWTC2}\cite{TOPC2} which 
can be the characteristics of this theory.

 In the SM, $t\bar{t}$ production at the Tevatron energy is dominated by the 
sub-process $q\bar{q}\rightarrow t\bar{t}$ \cite{SMsigma}. However, in a recent 
interesting paper, Eichten and Lane \cite{EL} showed that, in TC theories, 
color-octet technipions $\Pi^{0a}$ could make important contributions to 
$t\bar{t}$ production at the Tevatron via the gluon-gluon fusion sub-process 
$gg\rightarrow \Pi^{0a}\rightarrow t\bar{t}$ due to the large triangle-loop 
gluon-gluon-PGB coupling [cf. Fig. 1(a)], 
and such PGB could be tested by measuring the differential cross section
\footnote{The contribution of color-octet technirhos to the $t\bar{t}$ 
production has been considered in Refs.\cite{MWTC1}\cite{LS}\cite{HR}.}.
Considering the total $~t\bar{t}~$ production cross section, the color-singlet
technipion $~\Pi^0~$ also contributes. Furthermore, apart from the 
technifermion-loop contributions considered in Ref.\cite{EL} [fig. 1(a)], the 
isospin-singlet PBG's $\Pi^{0a}$ and $\Pi^0$ can also couple to the gluons 
through the top-quark triangle-loop \cite{SYZ}, and make contributions shown
in Fig.1(b). In the TOPCMTC theory, the top-pion $\Pi^0_t$, as an isospin-
triplet, can couple to the gluons through the top-quark triangle-loop in an 
isospin-violating way similar to the coupling of $\pi^0$ to the gluons in the 
Gross-Treiman-Wilczek formula \cite{GTW}, and the large isospin violation 
$~\frac {m_t-m_b}{m_t+m_b}\approx 1~$ makes its contribution to the 
$~t\bar{t}~$ production cross section important as well [cf. fig. 1(b)]. 
In this paper we study all these contributions to the production cross section 
of the sub-process $gg\rightarrow t\bar{t}$, and use the MRS set $A'$ parton 
distributions \cite{MRS} to calculate the cross sections at both 
the Tevatron and the LHC. The results of the total production cross sections 
show that, with these contributions, the cross section at the Tevatron is 
consistent with the new CDF datum for a certain range of the parameters,
and the new CDF datum does give constraints on the parameters in TOPCMTC 
models. The cross section at the $14~$TeV LHC is 
significantly larger than the SM prediction. The results of the differential 
cross sections show clear resonances of the PGB $\Pi^{0a}$ if its mass is in 
the reasonable range $400-500$~GeV. Therefore, this kind of model can be 
clearly tested by future experiments.

 This paper is organized as follows: Sec. 2 is devoted to the
calculation of the $gg\rightarrow t\bar{t}$ amplitude contributed by
the PGB's $\Pi^{0a}, \Pi^0,$ and $\Pi^0_t$. In Sec. 3, we present the
numerical results of the total contributions of $\Pi^{0a},~\Pi^0$ and 
$\Pi^0_t$ to the $t\bar{t}$ production cross sections at the Tevatron and 
the LHC in TOPCMTC models considering all fermion loops in Fig. 1(a)-(b).
The conclusions are given in Sec. 4.

\vspace{1cm}
\noindent
{\bf 2. The $gg\rightarrow t\bar{t}$ amplitude contributed by $\Pi^{0a},
\Pi^0$, and $\Pi^0_t$}

 In the topcolor-assisted multiscale technicolor theory, there are a lot
of PGB's. What are relevant to the $t\bar{t}$ production process are
the neutral technipions $\Pi^{0a}, \Pi^0$, and the neutral top-pion $\Pi^0_t$.
In the MWTC sector, the masses of $\Pi^{0a}$ and $\Pi^0$ have been estimated
to be $M_{\Pi^{0a}}\approx 200-600$~GeV and $M_{\Pi^0}\approx
100-300$~GeV, and the decay constants are $F=F_Q=F_L\approx
30-50$~GeV
\cite{MWTC1}. In the topcolor sector, if the topcolor scale is of the
order of $1$~TeV, the mass of $\Pi^0_t$ is around $~200$~GeV
and its decay constant is $F_t\approx 50$~GeV \cite{TOPC1}. Since these
PGB masses are not far from the $t\bar{t}$ threshold and $F$, $F_t$ are
all small, they may give important contributions to the $t\bar{t}$ 
production rates. In this section, we give the formulae for calculating
the production amplitudes $ gg\rightarrow \Pi^{0a}\rightarrow t\bar{t}$, 
$gg\rightarrow \Pi^0\rightarrow t\bar{t}$, and $gg\rightarrow \Pi^0_t
\rightarrow t\bar{t}$ shown in Fig. 1(a) and fig. 1(b).
These concern the couplings of the PGB's to fermions and to gluons, and
the PGB propagators.

    In the TOPCMTC theory, the top- and bottom-quark masses $~m_t~$ and
$~m_b~$ come from both the top-quark condensate and the ETC sector. It can 
be made that the large $~m_t~$ is mainly contributed by the top-quark
condensate, so that the ratio between the ETC contributed top- and
bottom-quark masses $~m'_t~$ and $~m'_b~$ is about the the same as the ratio
between the charm- and strange-quark masses, i.e. 
$~(m'_t/m'_b)\approx (m_c/m_s)\approx 10~$. This makes the value of the
electroweak parameter $~T~$ not too large in this theory. The value of 
$~m'_t~(m'_b)~$ depends on the parameters in the TOPCMTC model. For 
reaesonable values of the parameters, $~m'_t\sim 20-50~$GeV \cite{TOPC1}.

We first consider the couplings of the PGB's to $t\bar{t}$. At the
relevant energy scale, the PGB's can be described by local fields. In the
MWTC theory, the coupling of technipions to fermions are induced by ETC 
interactions and hence are model dependent. However, it has been generally 
argued that the couplings of the PGB's to the quark $q$ and antiquark 
$\bar{q}$ are proportional to $m'_q/F$ \cite{EL}\cite{RS}\cite{LS}, where 
$m'_q$ is the part of the quark mass acquired from the ETC. The PGB-$q$-
$\bar{q}$ vertices are of the following forms \cite{EL}\cite{LS}:

\begin{eqnarray}                                       %(1)
\frac{C_q m'_q}{\sqrt{2}F} \Pi^0(\bar{q} \gamma^5 q)\, ,  \hspace{2cm}
\frac{C_q m'_q}{F} \Pi^{0 a}(\bar{q} \gamma^5 \frac{\lambda^a}
{2} q )\, ,
\end{eqnarray}

\noindent
where $\lambda^a$ is the Gell-Mann matrix of the color group, $C_q$ is
a model dependent coupling constant which is expected to be typically
of $O(1)$ \cite{EL}\cite{RS}\cite{LS}. In the topcolor sector, by similar 
argument, we can obtain the interactions of the top-pions with the top and 
bottom quarks by replacing $m'_q$ by $m_q-m'_q$, and $F$ by $F_t$ in (1), i.e. 
\cite{TOPC2}

\begin{eqnarray}                                       %(2)
\frac{m_t - m'_t}{\sqrt{2}F_t} \bar{t} \gamma_5 t \Pi^0 +  
\frac{i}{\sqrt{2}}[ \bar{t} (1 - \gamma_5)b \Pi^+ +\frac{1}{\sqrt{2}}
 \bar{b} (1 + \gamma_5)t \Pi^-] \,,
\end{eqnarray}
\begin{eqnarray}                                       %(3)
\frac{m_b - m'_b}{\sqrt{2}F_t} \bar{b} \gamma_5 b \Pi^0  \,.
\end{eqnarray}

Next we consider the couplings of the PGB's to the gluons. Consider a general
formula for the coupling of a PGB to two gauge fields $B^\mu_1$ and $B^\nu_2$.
As far as the PGB's are described by local fields, the triangle fermion loops 
coupling the PGB's to $B_1$ and $B_2$ can be evaluated from the 
Adler-Bell-Jackiw anomaly. The general form of the effective PGB-$B_1$-$B_2$ 
interaction is \cite{DRK}\cite{LS}

\begin{eqnarray}                                       %(4)
\displaystyle{
\frac{1}{(1+ \delta_{B_1 B_2})}\left(
\frac{S_{\Pi B_1B_2}}{4\pi^2 F}\right) \Pi \epsilon_{\mu\nu\lambda\rho}
(\partial^\mu B^\nu_1)( \partial^\lambda B^\rho_2)}  \,,
\end{eqnarray}

\noindent
where $\Pi$ stands for $\Pi^0$, $\Pi^{0a}$ or $\Pi^0_t$; and when $B_1$ and 
$B_2$ are gluons, the factors $S_{\Pi gg}$ in different cases are as follows.

For $\Pi^0$ and $\Pi^{0a}$ with technifermion triangle-loop \cite{DRK},

\begin{eqnarray}                                       %(5)
S^{(Q,L)}_{\Pi^0 g_b g_c} = \sqrt{2}g_s^2 N_{TC} \delta_{bc}\,, \hspace{2cm}
S^{(Q,L)}_{\Pi^{0a} g_b g_c} = \sqrt{2}g_s^2 N_{TC} d_{abc}\,.
\end{eqnarray}

For $\Pi^0$ and $\Pi^{0a}$ with top-quark triangle-loop \cite{SYZ},

\begin{eqnarray}                                       %(6)
 S^{(t)}_{\Pi^0 g_bg_c} = \frac{C_t}{\sqrt {2}} 
g_s^2 J(R_{\Pi^0})\delta_{bc}\,, \hspace{2cm}
 S^{(t)}_{\Pi^{0 a} g_b g_c} =  \frac{C_t}{2}g_s^2 
d_{abc}J(R_{\Pi^{0a}})\,,
\end{eqnarray}

\noindent
with
\begin{eqnarray}                                 %(7)
J(R_{\Pi}) = - \frac{m'_t}{m_t}\frac{1}{R^2_{\Pi}}\int ^{1}_{0} \frac{dx}
{x(1 - x)} \ln[1 - R^2_{\Pi}x(1 -x)] \,,
\end{eqnarray}

\noindent
where $ R_{\Pi}\equiv  \frac{M_{\Pi}}{m_t} $.

 The coupling of $\Pi^0_t$ to gluons via the top-quark triangle-loop is 
isospin-violating  similar to the coupling of $\pi^0$ to gluons in the 
Gross-Treiman-Wilczek formula \cite{GTW}. It can also be calculated from the 
formula in Ref.\cite{SYZ} which gives \footnote{ \rm{It is proportional to 
the isospin-violating factor $~\frac{m_t-m_b}{m_t+m_b}\approx 1$.}}

\begin{eqnarray}                                       %(8)
 S_{\Pi^0_t g_b g_c} = \frac{1}{\sqrt 2} g_s^2 \delta_{bc} J(R_{\Pi^0_t}) \,,
\end{eqnarray}

\noindent
with
\begin{eqnarray}                                 %(9)
J(R_{\Pi^0_t}) = - \frac{m_t - m'_t}{m_t}\frac{1}{R^2_{\Pi^0_t}}\int ^{1}_{0} 
\frac{dx}{x(1 - x)} \ln[1 - R^2_{\Pi^0_t}x(1 -x)] \,,
\end{eqnarray}

\noindent
where $ R_{\Pi^0_t} \equiv \frac{M{\Pi^0_t}}{m_t} $.

 Finally the $\Pi$ ($\Pi^0, \Pi^{0a}, or \Pi^0_t$) propagator in Fig. 1 takes
the form

\begin{equation}                                 %(10)
\displaystyle \frac{i}{\hat{s}-M_\Pi^2+iM_\Pi\Gamma_\Pi}\,,
\end{equation}

\noindent
where $~\sqrt{\hat{s}}~$ is the c.m. energy and $~\Gamma_\Pi~$ is the total
width of the PGB $~\Pi$. The $~iM_\Pi\Gamma_\Pi~$ term in (11) is important
when $~\hat{s}~$ is close to $~M_\Pi^2$. The widths $~\Gamma_{\Pi^0}, 
~\Gamma_{\Pi^{0a}}~$, and $~\Gamma_{\Pi^0_t}~$ can be obtained as follows.
 
From (1) and (4) we see that the dominant decay modes of $~\Pi^0~$ are
$~\Pi^0\rightarrow b\bar{b}~$ and $~\Pi^0\rightarrow gg~$. So that

\begin{eqnarray}                                       %(11)
\Gamma_{\Pi^0}\approx \Gamma(\Pi^0 \rightarrow b \overline{b})+ 
 \Gamma(\Pi^0 \rightarrow g_a g_b) \,.
\end{eqnarray} 

\noindent
From (1) and (5), we can obtain

\begin{eqnarray}                                       %(12)
\Gamma(\Pi^0 \rightarrow  b \bar{b}) =\frac{3 C_b}{16\pi} \frac{m'^2_b 
M_{\Pi}}{F^2} \sqrt{1- \frac{4m^2_b}{M^2_{\Pi}}} \,,
\end{eqnarray} 
\noindent
and
\begin{eqnarray}                              %(13)                           
\Gamma(\Pi^0 \rightarrow g_a g_b) = \Gamma^{(Q,L)}(\Pi^0 \rightarrow g_a g_b )+
\Gamma^{(t)}(\Pi^0\rightarrow g_ag_b)=\Gamma^{(Q,L)}(\Pi^0\rightarrow g_ag_b)
\left|1 + \frac{C_t J(R_{\Pi^0})}{2 N_{TC}}\right|^2 
\nonumber
\end{eqnarray} 
\begin{eqnarray}                                   
=\frac{\alpha_s^2 N^2_{TC}}{16\pi^3} \frac{M^3_{\Pi}}{F^2} 
\left|1 + \frac{C_t J(R_{\Pi^0})}{2 N_{TC}}\right|^2 \,,
\end{eqnarray} 
where $\Gamma^{(Q,L)}$ and $\Gamma^{(t)}$ are the $~\Pi^0\rightarrow gg~$ rates
contributed by the technifermion loop and top-quark loop, respectively.

It has been shown \cite{SYZ}\cite{YLY} that $~\Pi^{0a}~$ decays dominantly 
into $~t\bar{t},~~gg~$, and $~gZ~$. So that

\begin{eqnarray}                                       %(14)
\Gamma_{\Pi^{0 a}} \approx \Gamma(\Pi^{0a} \rightarrow b \bar{b})+ 
 \Gamma(\Pi^{0 a}\rightarrow g_a g_b) + \Gamma(\Pi^{0 a} \rightarrow 
t \bar{t})+  \Gamma(\Pi^{0 a} \rightarrow g Z)\,.
\end{eqnarray} 
From (1), (4) and the value of $~S(\Pi^{0a} gZ)~$ given in Ref.\cite{RS}
\cite{LS},
we can obtain

\begin{eqnarray}                                       %(15)
\Gamma(\Pi^{0 a} \rightarrow  q \bar{q}) =\frac{C_q}{16\pi}
 \frac{m'^2_q M_{\Pi^{0a}}}{F^2} \sqrt{1- \frac{4m^2_q}{M^2_{\Pi^{0a}}}}\,,
 \hspace{1cm} q=t,b\,,
\end{eqnarray} 

\begin{eqnarray}                                       
\Gamma(\Pi^{0 a} \rightarrow  g_a g_b) =\Gamma^{(Q,L)}(\Pi^{0a}\rightarrow g_a 
g_b )\left|1 + \frac{C_t{ J(R_{\Pi^{0a}})}}{2\sqrt{2} N_{TC}}\right|^2  
\nonumber
\end{eqnarray} 
\begin{eqnarray}                                       %(16)
=\frac{5\alpha_s^2 N^2_{TC}}{384 \pi^3} \frac{M^3_{\Pi^{0a}}}{F^2} 
\left|1 + \frac{C_t {J(R_{\Pi^{0a}})}}{2\sqrt{2} N_{TC}}\right|^2 \,,
\end{eqnarray} 

\begin{eqnarray}                                       %(17)
\Gamma(\Pi^{0 a} \rightarrow  g Z) =\frac{\alpha \alpha_s}{144\pi^3} 
\left(\frac{N_{TC}}{4}\right)^2 {\tan}^2 \theta_W \frac{M^3_{\Pi^{0a}}}{F^2} 
\,.
\end{eqnarray} 

Since the top-pion mass is around $200~$GeV, it decays mainly into
$~b\bar{b}~$ and $~gg~$. Thus 

\begin{eqnarray}                                       %(18)
\Gamma_{\Pi^0_t}\approx \Gamma(\Pi_t^0 \rightarrow b \overline{b})+ 
 \Gamma_g(\Pi_t^0 \rightarrow g_a g_b) \,.
\end{eqnarray}
From (1) and (4) we obtain

\begin{eqnarray}                                       %(19)
\Gamma(\Pi_t^0 \rightarrow  b \overline{b}) =\frac{3 }{16\pi} \frac{(m_b - 
m'_b)^2}{F_t^2}M_{\Pi^0_t} \sqrt{1- \frac{4m^2_b}{M^2_{\Pi0_t}}} \,,
\end{eqnarray} 
and
\begin{eqnarray}                                       
\Gamma_g(\Pi_t^0 \rightarrow  g_a g_b)          %(20)
=\frac{\alpha_s^2 }{64\pi^3} \frac{M^3_{\Pi0_t}}{F_t^2}  |J(R_{\Pi^0_t})|^2\,.
\end{eqnarray} 

With the above formulae, we can obtain the following production amplitudes.

\begin{eqnarray}                                   %(21)
{\cal A}(g_bg_c\rightarrow \Pi^{0a}\rightarrow t\bar{t})=
{\cal A}^{(Q,L)}(g_bg_c\rightarrow \Pi^{0a}\rightarrow t\bar{t})+
{\cal A}^{(t)}(g_bg_c\rightarrow \Pi^{0a}\rightarrow t\bar{t})  \nonumber
\end{eqnarray}
\begin{eqnarray}
\displaystyle{= \frac {C_t m'_t g_s^2 [N_{TC}+C_tJ(R_{\Pi^{0a}})/(2\sqrt{2})]
 d_{abc}}
{4\pi^2\sqrt{2}F^2[\hat{s}-M_{\Pi^{0a}}+iM_{\Pi^{0a}}\Gamma_{\Pi^{0a}}]}
(\bar{t}\gamma_5 \frac{\lambda^a}{2} t)\epsilon_{\mu\nu\lambda\rho}
k^\mu_1 k^\rho_2 \epsilon^\nu_1 \epsilon^\lambda_2} \,,
\end{eqnarray}

\begin{eqnarray}                                 %(22)
{\cal A}(g_bg_c\rightarrow \Pi^0\rightarrow t\bar{t})=
{\cal A}^{(Q,L)}(g_bg_c\rightarrow \Pi^0\rightarrow t\bar{t})+
{\cal A}^{(t)}(g_bg_c\rightarrow \Pi^0\rightarrow t\bar{t})  \nonumber
\end{eqnarray}
\begin{eqnarray}
\displaystyle{=\frac{1}{\sqrt{2}} \frac{C_tm'_tg_s^2[N_{TC}+C_tJ(R_{\Pi^0})/2]
\delta_{bc}}{4\pi^2\sqrt{2}F^2[\hat{s}-M_{\Pi^0}+iM_{\Pi^0}\Gamma_{\Pi^0}]}
(\bar{t}\gamma_5 t)\epsilon_{\mu\nu\lambda\rho}
k^\mu_1 k^\rho_2 \epsilon^\nu_1 \epsilon^\lambda_2}\,,
\end{eqnarray}

and
\begin{eqnarray}                                %(23)
{\cal A}(g_bg_c\rightarrow \Pi^0_t\rightarrow t\bar{t})=
\displaystyle{
\frac{1}{2}\frac{(m_t-m'_t)g_s^2J(R_{\Pi^0_t})\delta_{bc}}
{8\pi^2F_t^2[\hat{s}-M^2_{\Pi^0_t}+
iM_{\Pi^0_t}\Gamma_{\Pi^0_t}]}(\bar{t}\gamma_5 t)
\epsilon_{\mu\nu\lambda\rho}k^\mu_1 k^\rho_2 \epsilon^\nu_1 
\epsilon^\lambda_2 }\,.
\end{eqnarray}

It is easy to obtain the SM tree-level $~t\bar{t}~$ production amplitudes

\begin{eqnarray}                              %(24)
\displaystyle{
{\cal A}^{SM}_{tree}(q\bar{q}\rightarrow t\bar{t})=\frac{
ig_s^2\bar{v}(p_{\bar{q}})\gamma^\mu\frac{\lambda^a}{2}u(p_q)
\bar{u}(p_t)\gamma_\mu\frac{\lambda^a}{2}v(p_{\bar{t}})}{\hat{s}}}\,,
\end{eqnarray}
and
\begin{eqnarray}                               %(25)
\displaystyle
{\cal A}^{SM}_{tree}(gg\rightarrow t\bar{t})={\cal
A}^{SM(s)}_{tree}(gg\rightarrow t\bar{t})+{\cal
A}^{SM(t)}_{tree}(gg\rightarrow t\bar{t})+{\cal
A}^{SM(u)}_{tree}(gg\rightarrow t\bar{t})\,,
\end{eqnarray}
with
\begin{eqnarray}                                
\displaystyle
{\cal A}^{SM(s)}_{tree}(gg\rightarrow t\bar{t})=
-ig_s^2[(k_2-k_1)^\mu(\epsilon_2\cdot\epsilon_1)+
(k_2+2k_1)\cdot\epsilon_2\epsilon_1^\mu
-(2k_2+k_1)\cdot\epsilon_1\epsilon_2^\mu] \nonumber
\end{eqnarray}
\begin{eqnarray}                                %(26)
\displaystyle
\hspace{-5cm} \times \frac{1}{\hat{s}}
\bar{u}(p_t)\gamma_\mu(if_{abc}\frac{\lambda^c}{2})v(p_{\bar{t}})\,,
\end{eqnarray}
\begin{eqnarray}                                %(27)
\displaystyle
{\cal A}^{SM(t)}_{tree}(gg\rightarrow t\bar{t})=-ig_s^2
\frac{\bar{u}(p_t)\rlap/{\epsilon}_1(\rlap/q-m_t)\rlap/{\epsilon}_2
\frac{\lambda^b}{2} \frac{\lambda^a}{2} v(p_{\bar{t}})}{q^2-m_t^2}\,,~~~~~~~~
q\equiv p_t-k_1\,,
\end{eqnarray}
\begin{eqnarray}                                 %(28)
\displaystyle
{\cal A}^{SM(u)}_{tree}(gg\rightarrow t\bar{t})=
{\cal A}^{SM(t)}_{tree}(gg\rightarrow t\bar{t})[1\leftrightarrow 2,\,
a\leftrightarrow b]\,,
%\frac{\bar{t}\rlap/{\epsilon}_1(\rlap/{q}'+m_t)\rlap{\epsilon}_2
%\frac{\lambda^a}{2} \frac{\lambda^b}{2} t}{\hat{u}-m_t^2}\,,
\end{eqnarray}
where $~k_1,~k_2~$ are the momenta of the two initial-state gluons,
%$~q\equiv p_t-k_2~$, $~q'\equiv p_t-k_1~$ (
$p_t$ is the momentum of the top-quark. 

Adding all these amplitudes together, we obtain the total 
$~t\bar{t}~$ production amplitude.

\vspace{1cm}
\noindent
{\bf 3. The $~t\bar{t}~$ production cross sections at the Tevatron and the
LHC}

Once we have the cross section at the parton level $~\hat{\sigma}~$,
the cross section at the hadron collider is obtained by convoluting
it with the parton distrbutions \cite{EHLQ}

\begin{eqnarray}                                 %(29)
 \sigma(pp(\bar{p})\rightarrow t\bar{t}) 
=\Sigma_{ij} \int dx_i dx_j f^{(p)}_i(x_i, Q) f^{(p(\bar{p}))}_j(x_j, Q) 
\hat{\sigma}(ij \rightarrow t \bar{t})\,,
\end{eqnarray}
where $i.j$ stand for the partons $~g, q~$ and $~\bar{q}~$; $x_i$ is the 
fraction of the longitudinal momentum of the proton (antiproton) carried by 
the $i$-th parton; $~Q^2\approx \hat{s}~$; and $~f^{(p(\bar{p}))}_i $ is the  
parton distribution functions in the proton (antiproton). In this paper, we 
take the MRS set $A'$ parton distrbution for $~f^{(p(\bar{p}))}_i$.
Taking into account of the QCD corrections, we shall multiply the obtained
$~\sigma~$ by a factor 1.5 \cite{SMsigma} as what was done in Ref.\cite{EL}.

The main purpose of Ref.\cite{EL} is to show the signal of $~\Pi^{0a}~$
at the Tevatron, so that they only calculated the technifermion-loop
contributions and neglected the interference between $~{\cal A}^{SM}_{tree}
(gg\rightarrow t\bar{t})~$ and $~{\cal A}(g_bg_c\rightarrow 
\Pi^{0a}\rightarrow t\bar{t})~$ as a first investigation. In this section, 
we present the cross sections at the Tevatron and the LHC in TOPCMTC 
models considering the contributions of $~\Pi^{0a}, \Pi^0~$ and $~\Pi^0_t~$ 
from FiG. 1(a) and Fig. 1(b) with the interferencess taken into account. 
In our calculation, we take the more updated parton distribution functions 
MRS set $A'$ instead of EHLQ set 1 taken in Ref.\cite{EL}. The fundamental 
SM parameters in our calculation are taken to be $~m_t=176~\rm{GeV},
~\sin^2\theta_W=0.231~$, and $~\alpha_s(\sqrt{\hat{s}})~$ the same as that in 
the MRS set $A'$ parton distributions. For the parameters in the TOPCMTC 
models, we simply take $~C_t=C_b=1$ and take the reasonable values 
$~F=40~$GeV, $~F_t=50~$GeV in this calculation. For the technipion masses, we 
fix $~M_{\Pi^0}=150~$GeV, and vary $~M_{\Pi^{0a}}~$ from $400$~GeV to 
$500$~GeV. The values of $~\Pi^0_t~$ and $~m'_t~$ depend on the parameters in 
the TOPCMTC models. To see how these values affect the cross sections, we 
take, some reasonable values for each of them, namely $~M_{\Pi^0_t}=150~ 
\rm{GeV~and}~350~$GeV, $~m'_t=20,~35~\rm{and}~50~$GeV.

 The results of the cross sections at the $1.8~$TeV Tevatron are listed in 
Table 1, in which $~\Delta\sigma^{(i)}_{t\bar{t}}~$ is the TOPCMTC correction
[ including the interferences between the TOPCMTC amplitudes (21)-(23) and 
the tree-level SM amplitudes (24)-(28) ] to the tree-level SM cross section 
in the total cross section $~\sigma^{(i)}_{t\bar{t}}~$, with 
$~i=1,2,3~$ corresponding to $~m'_t=20~\rm{GeV},~~35\rm{GeV},~~\rm{amd}~~50~
\rm{GeV}~$. We see that for most values of the parameters the cross sections 
$~\sigma_{t\bar{t}}~$ are consistent with the new CDF data except that the 
cross sections are too large for $~m_{\Pi^{0a}}=400~$GeV with $~m'_t\geq 35~
\rm{GeV}~$. Therefore the CDF data does give constraints on the values of 
$~m_{\Pi^{0a}}~~\rm{and}~~m'_t~$ which depends on the specific model. To see 
the constraints more precisely, we plot the cross section versus $~m'_t~$ in 
Fig. 2 (with $~m_{\Pi^0_t}=150~$GeV) and Fig. 3 (with $~m_{\Pi^0_t}=350~$GeV), 
in which the solid, dashed and, dotted lines stand for $~m_{\Pi^{0a}}=400,~
450~,~~\rm{and}~~500~$GeV, respectively. Comparing with the new CDF data (the
shaded band), we see that there are parameter ranges outside the band of the
CDF data, especially for $~m_{\Pi^0_t}=150~$GeV, the range of parameters
$~m_{\Pi^{0a}}=400~$GeV with $~m'_t>30~$GeV is disfavored; for
$~m_{\Pi^0_t}=350~$GeV, the range of parameters $~m_{\Pi^{0a}}=400~$GeV with
all $~m'_t>20~$GeV is disfavored.
 
In Figs. 4-6, we plot the differential cross sections 
$~\frac{d\sigma_{t\bar{t}}}{dm_{t\bar{t}}}~$ versus the $~t\bar{t}~$
invariant mass $~m_{t\bar{t}}~$ at the $~\sqrt{s}=1.8~$TeV Tevatron for 
various values of the parameters. We see that clear peak of the $~\Pi^{0a}~$ 
resonance emerges when $~m_{\Pi^{0a}}~$ lies in the range of 400 to 500 GeV. 
The larger the value of $~m'_t~$, the clearer the signal. This is because 
that the coupling in (1) is proportional to $~m'_t~$. For the case of 
$~m_{\Pi^0_t}=350~$GeV, the $~\Pi^0_t~$ peak can also be seen. Thus the model 
can be tested by the differential cross section for certain values of the 
parameters.

In Table 2, we list the values of $~\Delta\sigma_{t\bar{t}}~$ and 
$~\sigma_{t\bar{t}}~$ at the $~\sqrt{s}=14~$TeV LHC. We see that the
cross sections are much larger than those at the Tevatron due to the
fact that at the LHC $~t\bar{t}~$ production is dominated by gluon-fusion. 
The obtained cross section is significantly larger than the SM
predicted value, so that it can be easily tested by the future
experiment. In Figs. 7-9, we plot the differential cross sections at the LHC 
for various values of the parameters. We see that differential cross sections 
are similar to those at the Tevatron but the peaks are more significant
due to the same reason. So that the models can be better tested at the LHC.

\vspace{3cm}
{\bf 4.~Conclusions}

In this paper, we studied the $~t\bar{t}~$ production cross sections at the 
$~\sqrt{s}=1.8~$TeV Tevatron and the $~\sqrt{s}=14~$TeV LHC in the
TOPCMTC models. The TOPCMTC contributions are mainly via the $s$-channel
PGB's $~\Pi^{0a},~\Pi^0,~\rm{and}~\Pi^0_t~$ through gluon-fusion. We 
calculated both the diagrams in Fig.1(a) and Fig.1(b), and took into account 
the interferences between the tree level SM amplitudes [(24)-(28)] and the 
TOPCMTC amplitudes [(21)-(23)]. The MRS set $A'$ parton distribution functions 
are taken in this calculation. In the study, we take $~m_{\Pi^0}=150~$GeV and 
vary other parameters in the models. Our results show that the production 
cross sections are enhanced by the TOPCMTC contributions. The present CDF 
datum on the production cross section gives constraints on the model-dependent 
parameters $~m_{\Pi^{0a}}~\rm{and}~m'_t~$, i.e. $~m_{\Pi^{0a}}=400~$GeV
with large $~m'_t~$ is disfavored. In the differential cross sections, clear 
peaks of the $~\Pi^{0a}~$ and $~\Pi^0_t~$ can be seen for reasonable range of 
the parameters, so that the models are experimentally testable at the 
Tevatron and the LHC. The cross section at the LHC is significantly
larger than the SM predicted value, and the peaks are more significant at the 
LHC than at the Tevatron due to the fact that $~t\bar{t}~$ production at the 
LHC is dominated by gluon-fusion. Therefore the models can be better tested 
at the LHC.

\vspace{1cm}
\noindent {\bf ACKNOWLEDGMENT} 

  C.-X. Yue would like to thank B.-L. Young for his valuable discussions. 
This work is supported by National Natural Science Foundation of China,
the Natural Foundation of Henan Scientific Committee, and the Fundamental
Research Foundation of Tsinghua University.

%Tevatron tree level $\sigma_0=3.168(q\bar q)+0.239(gg)=3.407(pb)$

%LHC tree level $\sigma_0=0.081(q\bar q)+0.412(gg)=0.493(nb)$

\newpage

 \begin{table}
%\caption{ }
{\bf Table 1.}  $t\bar{t}~$  production cross section $~\sigma(g g 
\rightarrow \pi^{0(a)}(\pi^0,\pi^0_t) \rightarrow t \bar{t})~$ at the 
$~\sqrt{s} =1800$~GeV Tevatron in the topcolor-assisted multiscale walking 
technicolor model with $~m_{\pi^0}=150~$GeV. $~\Delta\sigma^{(i)}~$ is the
TOPCMTC correction to the tree-level SM cross section and $~\sigma^{(i))}_
{t\bar{t}}~$ is the total cross section, where $~i=1,2,3~$ correspond to 
$~m'_t=20~\rm{GeV},~~35~\rm{GeV},~~\rm{and}~~50~\rm{GeV}~$, respectively.  
A factor 1.5 of QCD corrections has been taken into account.
 
%\bigskip

\begin{center}
\begin{tabular}{||c|c|c|c|c|c|c|c||}
\hline
 $M_{\pi_t}$(GeV) & $M_{\pi^{0 a}}$(GeV)  &  $\Delta\sigma^{(1)}$ (pb) & 
 $\sigma^{(1)}_{t\bar t}$(pb) & $\Delta\sigma^{(2)}$ (pb) & 
 $\sigma^{(2)}_{t\bar t}$ (pb) &$\Delta\sigma^{(3)}$ (pb) & 
 $\sigma^{(3)}_{t\bar t} $ (pb)\\ \hline
 150 & 400 &2.750 & 7.861 &  5.678 & 10.789 & 7.829 & 12.940\\ \hline 
 150 & 450 & 1.350 & 6.461 &  2.688 & 7.798 & 3.618 & 8.729\\ \hline   
 150 & 500 & 0.632 & 5.743 &  1.245 & 6.356 & 1.680 & 6.791\\ \hline\hline
 350 & 400 & 4.638 & 9.749 &  7.127 & 12.238 &  8.834 & 13.945\\ \hline   
 350 & 450 & 2.970 & 8.081 &  3.965 & 9.076 & 4.571& 9.682\\ \hline 
 350 & 500 & 2.279 & 7.390 &  2.480 & 7.591 & 2.598 & 7.709\\ \hline
\end{tabular}
\end{center}
\bigskip
%\caption{ }
{\bf Table 2.} $~t\bar{t}~$  production cross section $~\sigma(g g 
\rightarrow \pi^{0(a)}(\pi^0,\pi^0_t) \rightarrow t \bar{t})~$ at the  
$\sqrt{s} =14$~TeV LHC in the topcolor-assisted multiscale walking technicolor 
model with $~m_{\pi^0}=150~$GeV. $~\Delta\sigma^{(i)}~$ is the TOPCMTC
correction to the tree-level SM cross section and $~\sigma^{(i)}_{t\bar{t}}~$ 
is the total cross section, where $~i=1,2~$ correspond to $~m'_t =20~\rm{GeV}
~~\rm{and}~~35~\rm{GeV}~$, respectively. A factor 1.5 of QCD corrections has 
been taken into account.

%\bigskip

\begin{center}
\begin{tabular}{||c|c|c|c|c|c||}
\hline
  $M_{\pi_t}$(GeV) & {$ M_{\pi^{0 a}}$(GeV) } & {$\Delta\sigma^{(1)} $(nb)} &
 $\sigma^{(1)}_{t\bar t}$(nb) &  
 $\Delta\sigma^{(2)}$(nb) &  $\sigma^{(2)}_{t\bar t} $(nb)\\ \hline
 150 & {\tt 400} &  2.753 & 3.493 & 5.577 & 6.317 \\ \hline 
 150 & {\tt 450} &  2.073 & 2.813 & 4.167 & 4.907 \\ \hline   
 150 & {\tt 500} &  1.596 & 2.336 & 3.180 & 3.920 \\ \hline \hline
 350 & {\tt 400} &  3.791 & 4.531 & 6.293 & 7.033 \\ \hline   
 350 & {\tt 450} &  3.182 & 3.922 & 5.006 & 5.746 \\ \hline 
 350 & {\tt 500} &  2.568 & 3.308 & 3.978 & 4.718\\ \hline  
\end{tabular}
\end{center}
\end{table}

\newpage
\begin{center}
{\Large \bf Figure Captions}\\
\end{center}
\vspace{0.6cm}
{\bf Fig. 1.} Feynman diagrams for the TOPCMTC contributions to the 
$~t\bar{t}~$ productions at the Tevatron and the LHC.\\
{\bf (a)}. Techniquark loop contributions.\hspace{0.5cm}
{\bf (b)}. Top-quark loop contributions.\\ \\
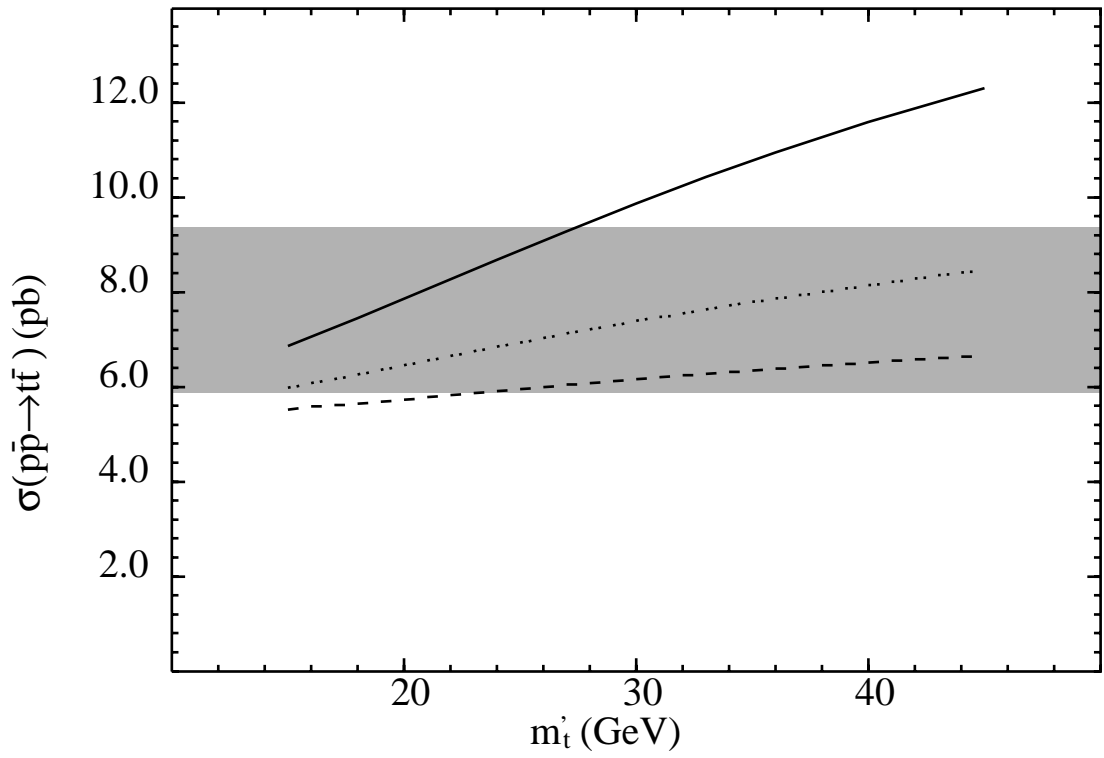
{\bf Fig. 2.} The plot of $~\sigma_{t\bar{t}}~$ versus $~m'_t~$ for 
$~m_{\Pi^0_t}=150~$GeV at the Tevatron. The solid, dashed, and dotted lines 
stand for $~m_{\Pi^{0a}}=400,~450,~\rm{and}~500~$GeV, respectively. The
CDF data is indicated by the shaded band.\\ \\
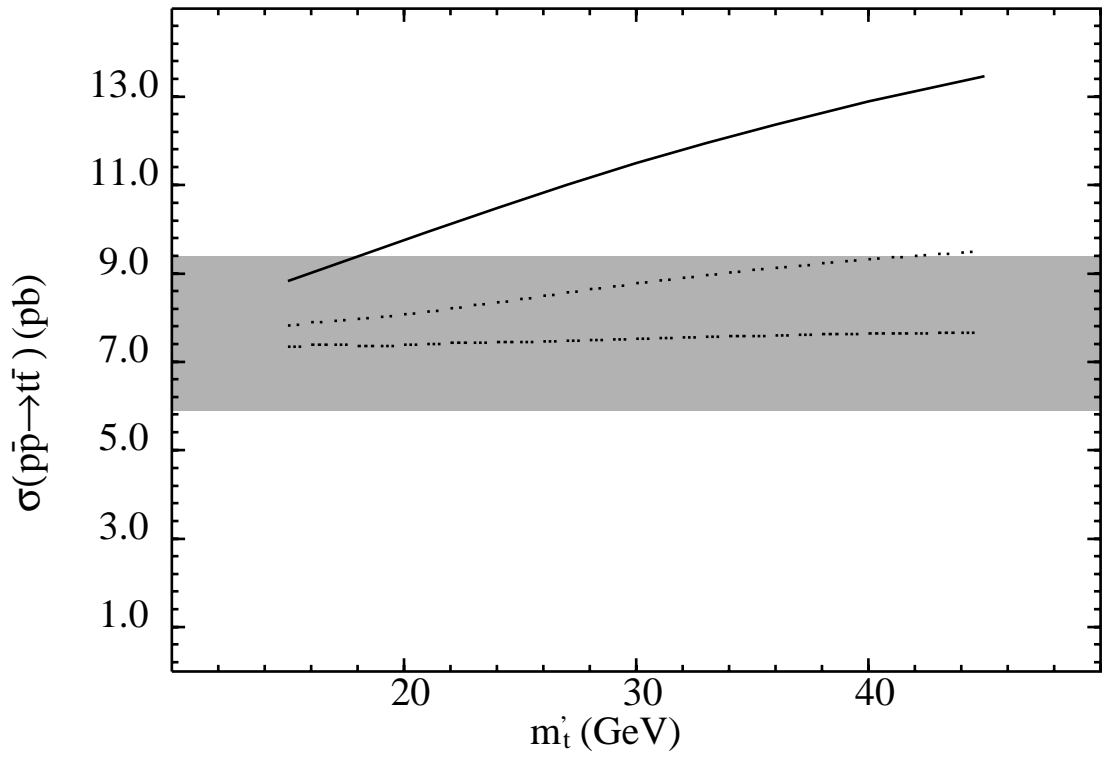
{\bf Fig. 3.} Same as {\bf Fig. 2} but for $~m_{\Pi^0_t}=350~$GeV.\\ \\
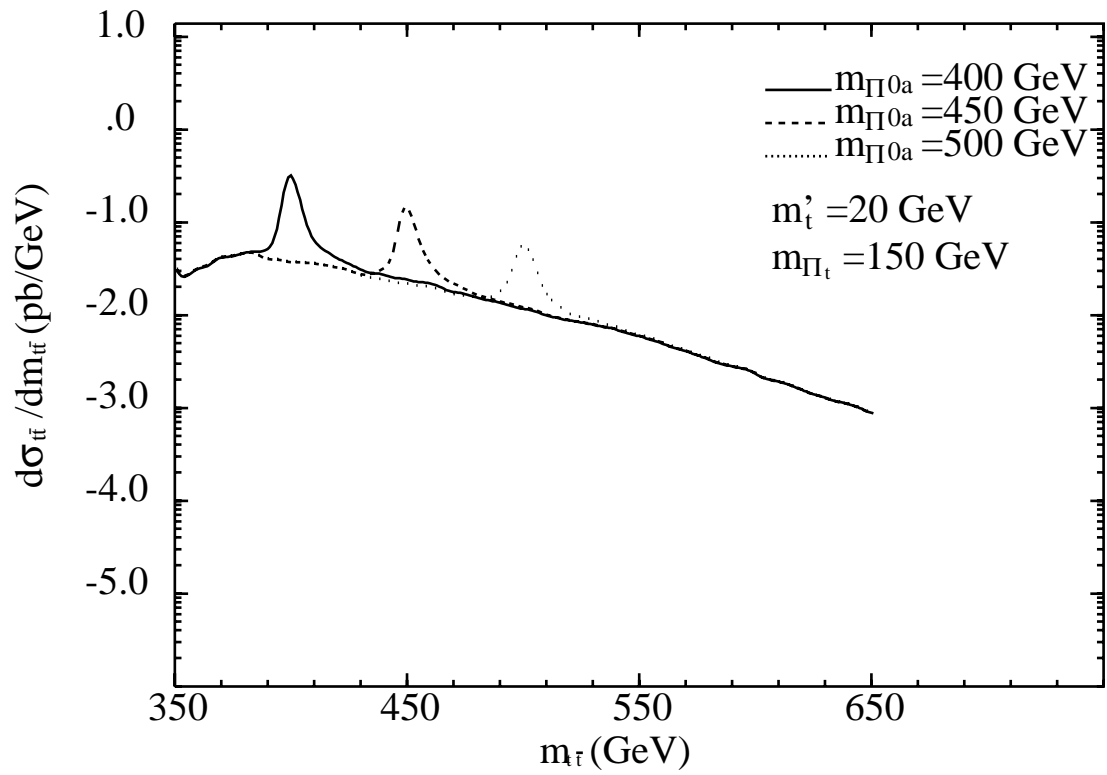
{\bf Fig. 4.} Differential cross section $~\frac{d\sigma_{t\bar{t}}}
{dm_{t\bar{t}}}~$ (in logarithmis scale) versus the $~t\bar{t}~$ invariant 
mass $~m_{t\bar{t}}~$ at the Tevatron for $~m_{\Pi^{0a}}=400,~450,~\rm{and}
~500~$GeV with $~m_{\Pi^0}=150~$GeV, $~m'_t=20~$GeV, and $~m_{\Pi^0_t}=
150~$GeV.\\ \\
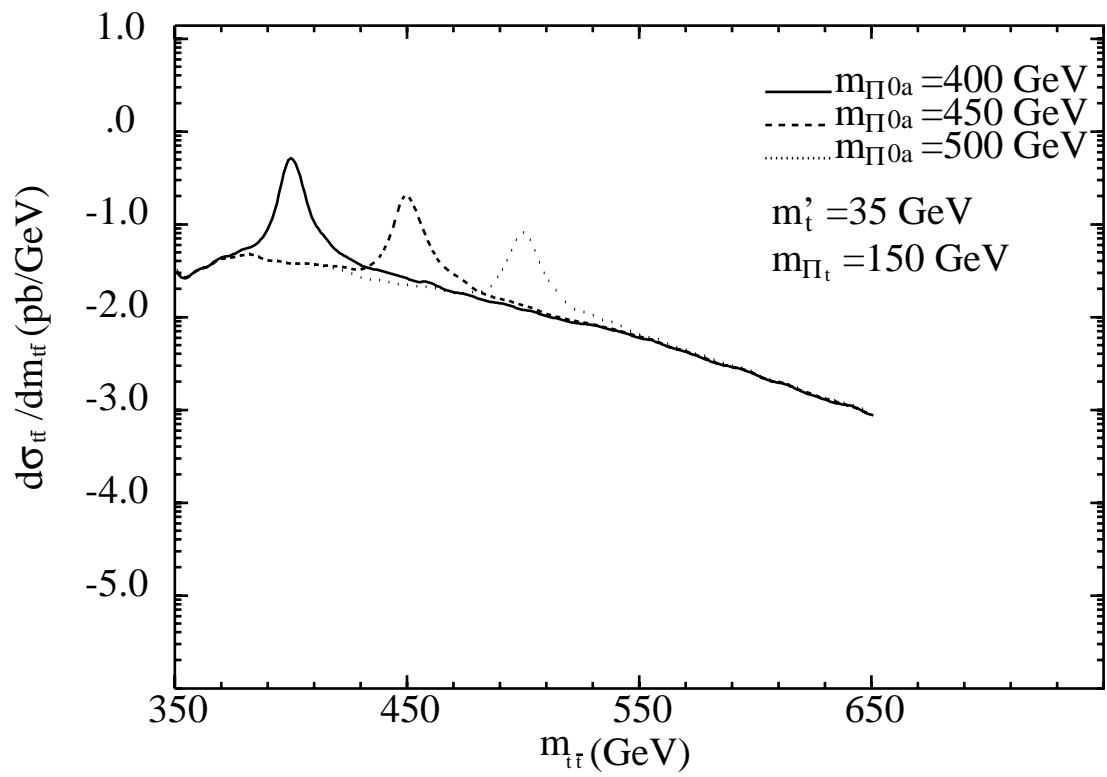
{\bf Fig. 5.} Same as {\bf Fig. 3.} but with $~m'_t=35~$GeV and 
$~m_{\Pi^0_t}=150`$GeV.\\ \\
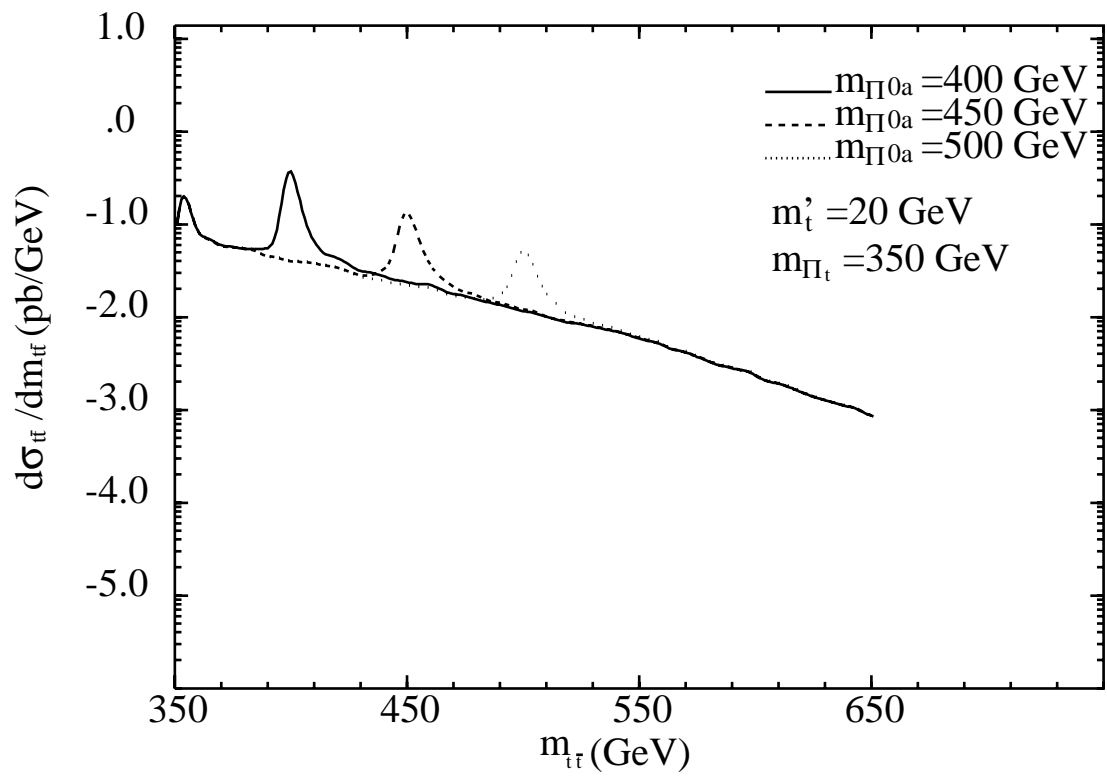
{\bf Fig. 6.} Same as {\bf Fig. 3.} but with $~m'_t=20~$GeV and $~m_{\Pi^0_t}
=350~$GeV.\\ \\
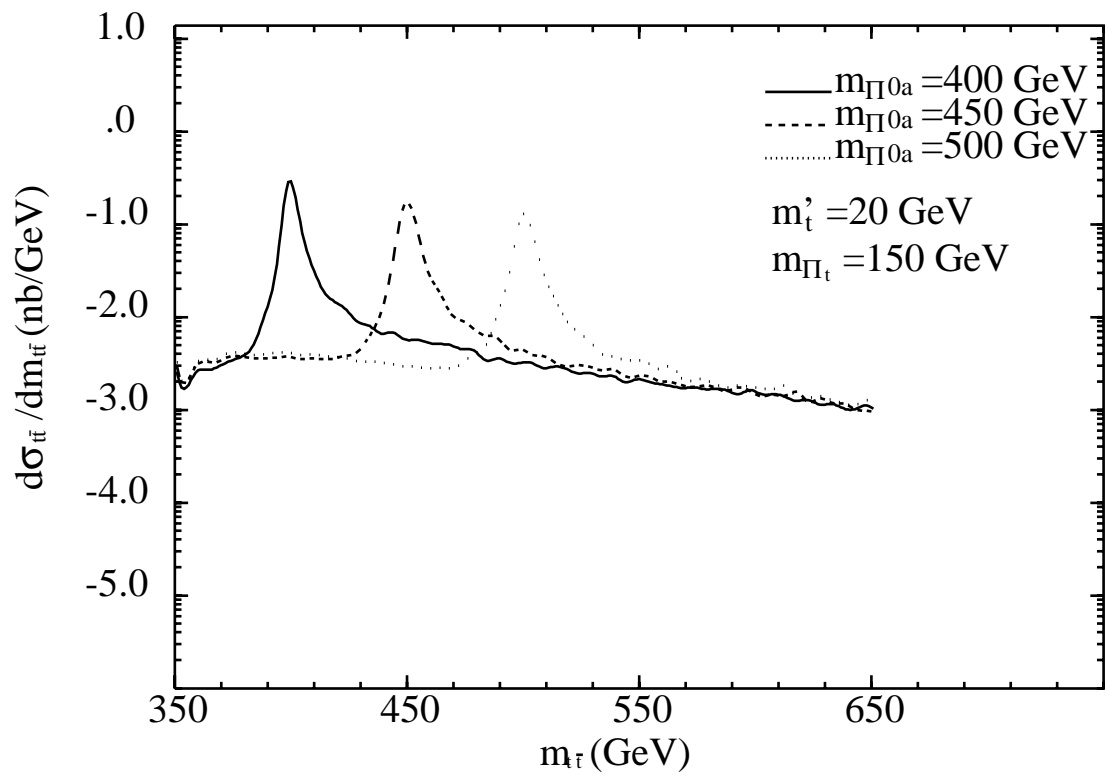
{\bf Fig. 7.} Same as {\bf Fig 3.} but at the LHC.\\ \\
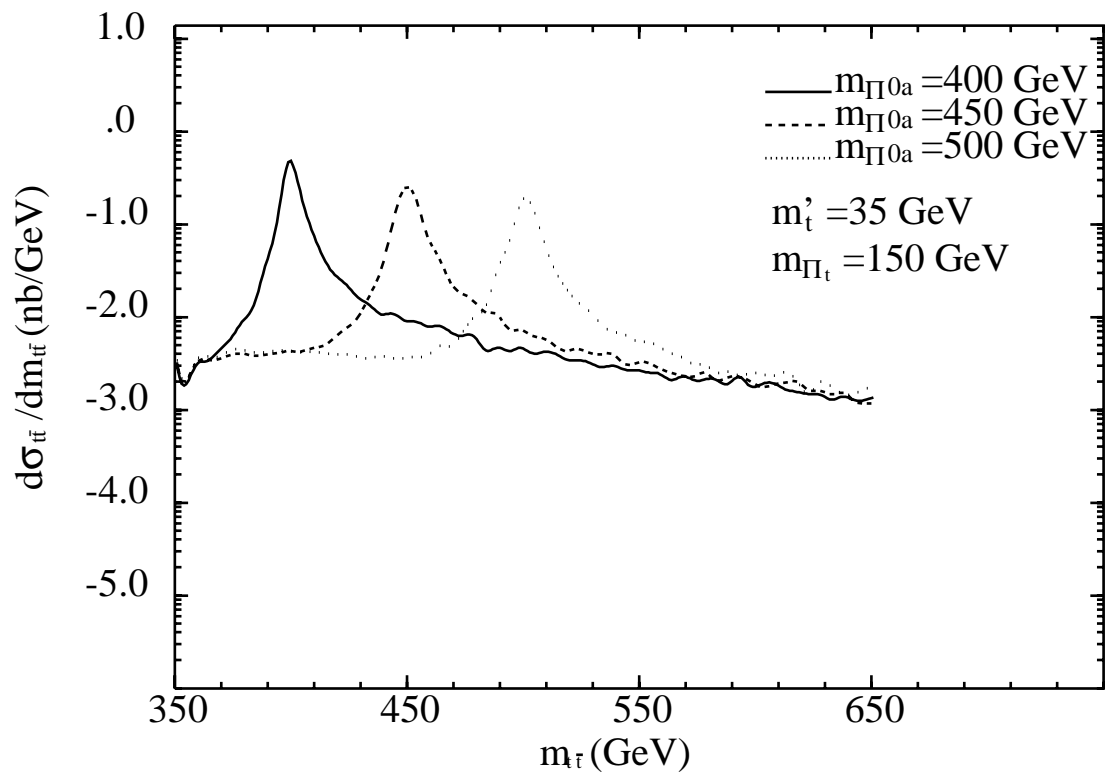
{\bf Fig. 8.} Same as {\bf Fig. 4} but at the LHC.\\ \\
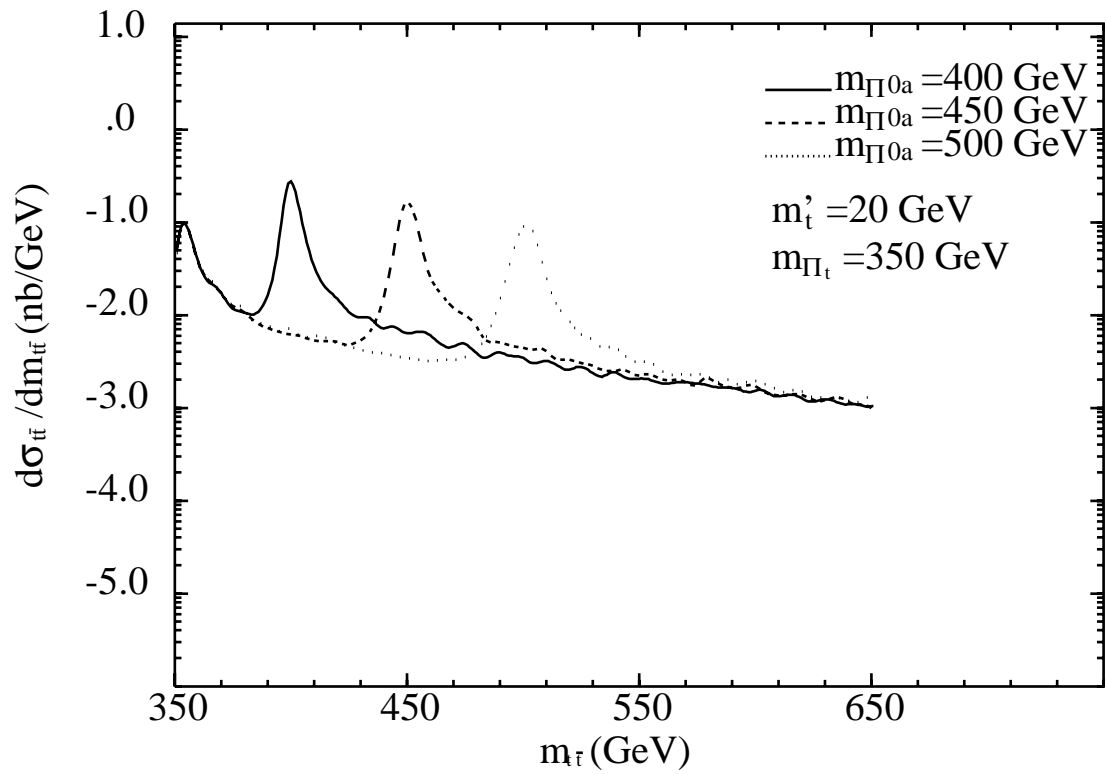
{\bf Fig. 9.} Same as {\bf Fig. 5} but at the LHC.

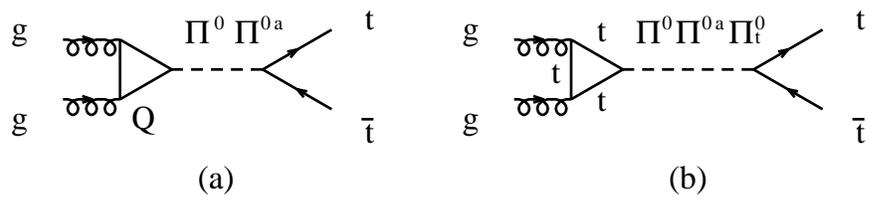
\end{document}